\documentclass{article18}
\usepackage{filecontents}
\usepackage{mathtools}
\usepackage{amsmath}
\usepackage[linesnumbered,ruled]{algorithm2e} 
\usepackage[utf8]{inputenc}
\usepackage[english]{babel}
\usepackage{hyperref}

\DeclarePairedDelimiter\abs{\lvert}{\rvert}
\DeclarePairedDelimiter\norm{\lVert}{\rVert}
\makeatletter
\let\oldabs\abs
\def\abs{\@ifstar{\oldabs}{\oldabs*}}
\let\oldnorm\norm
\def\norm{\@ifstar{\oldnorm}{\oldnorm*}}
\makeatother

\title{A New Optimal Algorithm for Computing the Visibility Area of a simple Polygon from a Viewpoint}
\author{Hamid Hoorfar \thanks{Department of Computer Engineering and Information Technology, Amirkabir University of Technology (Tehran Polytechnic), {\tt \{hoorfar,ar\_bagheri\}@aut.ac.ir}}\and Alireza Bagheri\footnotemark[1]}
\makeatletter
\let\runauthor\@author
\let\runtitle\@title
\makeatother
\begin{document}
\maketitle
\begin{abstract} 
Given a simple polygon $ \mathcal {P} $ of $ n $ vertices in the Plane. We study the problem of computing the visibility area from a given viewpoint $ q $ inside $ \mathcal {P} $ where only sub-linear variables are allowed for working space.
Without any memory-constrained, this problem was previously solved in $ O(n) $-time and $ O(n) $-variables space. In a newer research, the visibility area of a point be computed in 
$ O(n) $-time, using $ O(\sqrt{n}) $ variables for working space. In this paper, we present an optimal-time algorithm, using $ O(c/\log n) $ variables space for computing visibility area, where $ c<n $ is the number of critical vertices.
We keep the algorithm in the linear-time and reduce space as much as possible.
\end{abstract}

\section{Introduction}
The visibility area of a simple polygon $ \mathcal {P} $ from an internal viewpoint $ q $ is the set of all points of $ \mathcal {P} $ that are visible from $ q $, since two points $ p $ and $ q $ are visible from each other whenever the line-segment $ pq $ does not intersects $ \mathcal {P} $ in more than one connected component i.e. the visibility area of viewpoint $ q $ inside $ \mathcal{P} $ is the set of all points in $ \mathcal{P} $ that are visible from $ q $. In similar definition, we say that the points $ p $ and $ q $ inside $ \mathcal{P} $ are $ k $-visible from each others if the line segment $ pq $ intersects the boundary of $ \mathcal{P} $ at most in $ k $ times. The $ k $-visibility area of viewpoint $ q $ inside $ \mathcal{P} $ is the set of all points in $ \mathcal{P} $ that are $ k $-visible from $ q $.
One of the important problem studies in computational geometry is visibility~\cite{ghosh2007computing}. Joe and Simpson presented the first correct optimal algorithm for computing visibility area (region) from a point in linear-time and linear-space~\cite{joe1987corrections,toth2017handbook}. In constrained-memory model, some algorithms has restricted memory (generally, not more than logarithmic in the size of input and typically constant). In constant-memory, algorithms do not have a long-term memory structure. Processing and deciding of current data must be at the moment. Using constant space, there is not enough memory for storing previously seen data or even for the index of this data. However, in the constraint-memory mode, there is $ O(s) $ bits working space during running algorithm, certainly, less than  $ O(n) $. Several researchers worked on constrained-memory algorithms in visibility~\cite{abrahamsen2013constant,asano2013memory,barba2011computing,de2012space}. Some recent related works are reviewed here. Computing visibility area of a viewpoint $ q $ in the simple polygon $ \mathcal {P} $ with extra  $ O(1) $ variables ($ O(\log n)$ bits per variable) has time-complexity of $ O(n\overline{r})$ where $ \overline{r} $ is the cardinality of the reflex vertices of $ \mathcal {P} $ in the output ~\cite{barba2014computing} while there exists a linear-time algorithm for computing the visibility area without constrained memory, using $ O(n) $ variables~\cite{asano1986visibility}. If there is $ O(s) $ variables for the computing, the time complexity of computing visibility area decreases to $ O(\frac{n\overline{r}}{2^{s}}+n \log \overline{r}) $ where $ s \in O(\log \overline{r}) $. It was presented a time-space trade-off algorithm which reports the $ k $-visibility area for $ q  \in {\mathcal {P}}$ in $ O(\frac{cn}{s} + n \log s + \min \{\frac{kn}{s}, n \log \log_s n\}) $ time, the algorithm uses $ O(s) $ words with $ O(\log n) $ bits each of workspace and $ c \leq n $ is the number of vertices of $ \mathcal {P} $ where the visibility area changes for the viewpoint $ q $~\cite{bahoo2017time}. If $ k=1 $ and $ s=1 $, their algorithm time complexity is $ O(cn) $, in fact, using $ O(\log n) $ bits space. Also, it is presented a linear-time algorithm for computing visibility area, using $O\left(\sqrt{n}\right)$ variables~\cite{de2012space}.

\textbf{\textit{Our General Result:}}\\
We provide an optimal-time algorithm of $ O(n) $ time, and $ O(c+\log n) $ bits space (or $ O({c}/{\log n}) $ variables of $O(\log n)  $ bits each) for computing visibility area of a viewpoint $ q $ inside $ \mathcal {P} $, where $ c $ is the number of critical vertices in $ \mathcal{P} $. Also, $ c\leq r<n $, where $ r $ is the number of reflex vertices of $ \mathcal{P} $. We keep the algorithm linear-time and reduce space as much as possible.
\newpage
\section{Preliminaries}
\label{ss:ss1}
We present our algorithm in the sub-linear space. Assume that a simple polygon $ \mathcal P $ is given in a read-only array $ \mathcal{A} $ of $ n $ vertices (each element is wrote in $ \log(n) $ bits) in counterclockwise order along the boundary and there is a read-only variable that is contained a query viewpoint $ q $ inside $ \mathcal{P} $, also, it has $ \log(n) $ bits. It is called \textit{input} with random access. We can use $ O(s) $ writable and readable variables of size $ O(\log n) $ as \textit{workspace} of the algorithm. The workspace is both writable and readable during the running of the algorithm. In this model, the \textit{output} is available in a write-only array that will contain the boundary of visibility area of $ q $ in counterclockwise order, at the end of the process. 
\begin{figure}[t]
    \centering
    \includegraphics[width=0.5\textwidth]{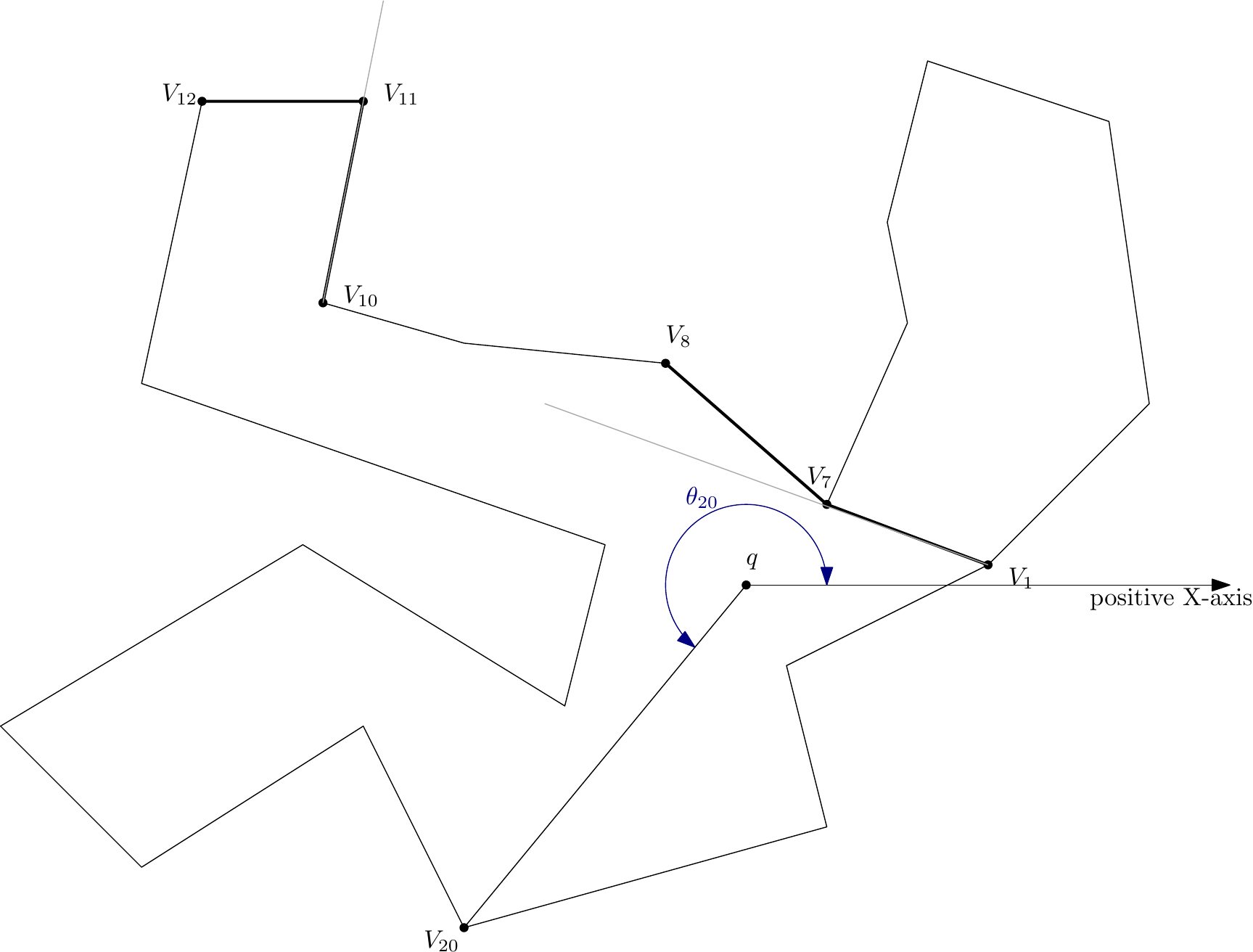}
    \caption{Illustration of the angle of $ p_i $, the angle, $ \theta_{20} $ is the angle of $ qp_{20} $, the $ \angle{V_1V_7V_8} $ is a clockwise turn and the angle $ \angle{V_{10}V_{11}V_{12}} $ is a counterclockwise turn.}
    \label{fi:fig1}
\end{figure}
For every vertex $ p_i $ of $ \mathcal P $, we compute the angle between vector(ray) $ \overrightarrow{qp_i} $ and the positive $ X $-axis as denoted by $ \theta_i $ (radian). For every $ \theta_i $, where $ i $ between $ 0 $ and $ n-1 $, $ 0\leq \theta_i\leq 2\pi $. If $ p_i $ is a vertex, then $ \rho_i $ equals the length of $ qp_i $.
For every $ 3 $ vertices $ v_1 $,$ v_2 $ and $ v_3 $, the angle $ \angle{v_1v_2v_3} $ is called \textit{counterclockwise turn}, if and only if $ v_3 $ is placed on the left side of the vector $ \overrightarrow{v_1v_2} $, otherwise, if and only if $ v_3 $ is placed on the right side, it is called \textit{clockwise turn}, see Figure~\ref{fi:fig1}. Sometimes, for simplicity we denote angle $ \angle{v_1v_2v_3} $ as $ \angle{v_2} $. For two vertices $ \nu $ and $ \nu' $, let $ \Gamma_{\nu} $ is the line which is contained $ \nu $ and $ q $, let $ e_1 $ and $ e_2 $ are two edges which are contained $ \nu' $, so, the intersection of $ e_1 $ and $ \Gamma_{\nu} $, or $ e_2 $ and $ \Gamma_{\nu} $, that one touches the polygon from internal of $ \mathcal{P} $(from direction of $ q $), is called shadow of $ \nu $, we denote it by $ S(\nu) $ or $ \overline{\nu} $, see Figure~\ref{fi:fig2}(A) and~\ref{fi:fig2}(B).
\begin{figure}
    \centering
    \includegraphics[width=0.8\textwidth]{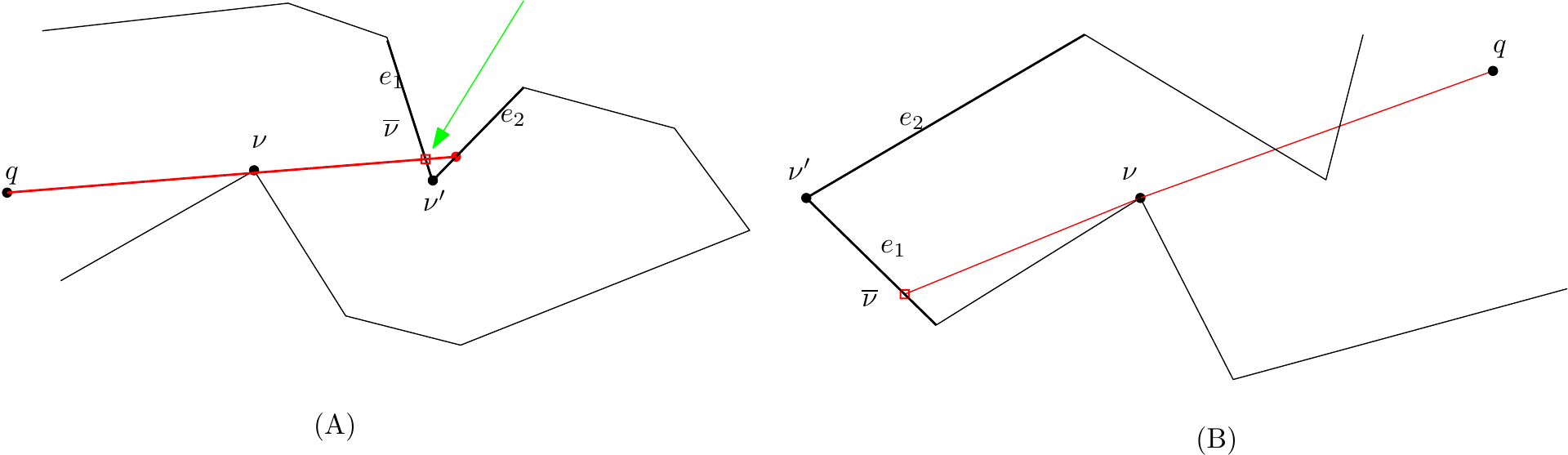}
    \caption{(A)the red box is shadow of $ \nu $, in this case, there are two intersections between $ e_1 $, $ e_2 $ and $ q\nu $, so, the red circle is not the shadow. (B)the red box is shadow of $ \nu $, in this case, there is  one intersection between $ e_1 $, $ e_2 $ and $ q\nu $.}
    \label{fi:fig2}
\end{figure}
If $ \theta_{c-1}<\theta_c >\theta_{c+1} $ and $ \angle p_c $ is a clockwise turn, then $ p_c $ is called \textit{critical-max vertex} and if $ \theta_{c-1}>\theta_c <\theta_{c+1} $ and $ \angle p_c $ is a counterclockwise turn, then $ p_c $ is called \textit{critical-min vertex}. If a vertex is critical-max or critical-min, it is called \textit{critical}.

\section{An algorithm for finding effective critical vertices of simple polygons}
In this section, the aim is presenting a linear-time algorithm for finding \textit{effective} critical vertices of $ \mathcal{P} $ in the sub-linear space model. The effective critical vertex is a critical vertex that is visible from $ q $. We use $ O(c+\log n) $ bits for working space of our algorithm to marking positions of effective critical vertices among all critical vertices of $ \mathcal{P} $ where $ c<n $ is the number of critical vertices, therefore, variable $ i $ in Algorithm~\ref{al:algo1} is an index for the critical vertices. This workspace equals to $ O(c/\log n) $ variables that have $ \log n $ bits each. Assume array $ W $ with $ r $ bits is a part of the algorithm workspace and we can also use a constant number of variables during computing. If $ k $-th bit of $ W $ (denoted by $ w_k $) is $ 1 $, then it means the $ k $-th critical vertex of the boundary is visible from $ q $ (effective), else the $ k $-th critical vertex of the boundary is not effective. Using array $ W $, we can save the position of effective critical vertices among all critical vertices. In the following, we will explain how our algorithm works.\\
At the first, all elements of $ W $ have been set to $ 1 $. We traverse among the boundary in both directions, clockwise and counterclockwise order. If the edges of a vertex (generally, critical vertex) causes $ j $-th critical vertex not to be visible from $ q $, we set $ w_j=0 $ that means it is not effective. See Algorithm~\ref{al:algo1}.
\begin{algorithm}[h]
    \KwIn{Vertices of simple polygon $ \mathcal{P} $ in random-access array $ \mathcal{A} $ and a view point $ q $ inside $ \mathcal{P} $}
    \KwOut{Critical vertices of $ \mathcal{P} $ that are not effective.}
    Set $ V_0,V_1,V_2 $ three vertices with the smallest angle in $ \mathcal{A} $, respectively\;
    $ i\leftarrow$the number of critical vertices in $ \{V_0,V_1,V_2\} $\;
    \While{$not( \theta_0<\theta_1>\theta_2 $ and $ \angle V_0V_1V_2 $ is clockwise turn)}{
    $ V_0\leftarrow V_1 $; $ V_1\leftarrow V_2 $; $ V_2\leftarrow $next-vertex\;
    \If{$ V_2 $ is critical}{$ i\leftarrow i+1 $;}
    }
    \While{$ \theta_1>\theta_2$}{
	$ V_2\leftarrow $next-vertex\;
	\If{$ V_2 $ is critical}{$ i\leftarrow i+1 $; $ w_i\leftarrow 0 $;}
	}
    Goto 3\;
    Set $ V_0,V_1,V_2 $ three vertices with the largest angle in $ \mathcal{A} $, respectively\;
	\While{$not( \theta_0>\theta_1<\theta_2 $ and $ \angle V_0V_1V_2 $ is counterclockwise turn)}{
    $ V_0\leftarrow V_1 $; $ V_1\leftarrow V_2 $; $ V_2\leftarrow $previous-vertex\;
   	\If{$ V_2 $ is critical}{$ i\leftarrow i-1$;}
    }
    \While{$ \theta_1<\theta_2$}{
	$ V_2\leftarrow $previous-vertex\;
	\If{$ V_2 $ is critical}{$ i\leftarrow i-1 $; $ w_i\leftarrow 0 $;}
	}
    Goto 17\;
\caption{Computing the effective critical vertices.}
\label{al:algo1}
\end{algorithm}\\
After running Algorithm~\ref{al:algo1}, the angles sequence of critical-min vertices ordered counterclockwise will be ascending. Also, the angles sequence of critical-max vertices ordered counterclockwise will be ascending, too. The running of this algorithm may not lead to the angles sequence of all critical vertices is ascending, ordered counterclockwise. So, we use Algorithm~\ref{al:algo2} to merge these two ascending sequences to obtain the sequence of all effective critical vertices.
\begin{algorithm}[h]
    \KwIn{Positions of effective critical vertices, flagged in array $ W $, ordered counterclockwise.}
    \KwOut{Ascending sequence of all effective critical vertices, flagged in array $ W $.}
    Set $ V_0,V_1,V_2 $ three critical vertices with the smallest angle in $ \mathcal{A} $, that are flagged by $ W $, respectively; $ i\leftarrow$the number of critical vertices from the beginning to now\;
    \While{$not( \theta_0>\theta_1 and \theta_0>\theta_2 $)}{
    $ V_0\leftarrow V_1 $; $ V_1\leftarrow V_2 $; $ V_2\leftarrow $next-critical-vertex with $ w_{i+1}=1 $\;
    Increase $ i$ by one\;
    }
    $ w_{i-1}\leftarrow 0 $\;$V_1 \leftarrow V_2 $; $ V_2\leftarrow $next-critical-vertex with $ w_{i+1}=1 $\;
	Increase $ i $ by one\;
	Goto 2\;
    Set $ V_0,V_1,V_2 $ three critical-vertices with the largest angle in $ \mathcal{A} $, respectively\;
	\While{$\theta_0>\theta_1 $}{
	    $ V_0\leftarrow V_1 $; $ V_1\leftarrow V_2 $; $ V_2\leftarrow $previous-critical-vertex with $ w_{i-1}=1 $\;
	    Decrease $ i$ by one\;
	    }
	    \While{$\theta_0<\theta_2$}{
	     $ w_{i+1}\leftarrow 0 $\;$V_1 \leftarrow V_2 $; $ V_2\leftarrow $previous-critical-vertex with $ w_{i-1}=1 $\;
	     Decrease $ i $ by one\;
	    }
	    Set $ \rho\leftarrow $ shadow of $ p_0 $ on $ p_1 $\;
	    \eIf{$ q\rho > qp_0$}{
	    $ w_{i+1}\leftarrow 0 $\;
	    }{
	    $ w_{i+2}\leftarrow 0 $;
	    $V_0 \leftarrow V_1 $;
	    }
	    $V_1 \leftarrow V_2 $; $ V_2\leftarrow $previous-critical-vertex with $ w_{i-1}=1 $\;
	    Decrease $ i $ by one\;
	    Goto 11\;
\caption{Modifying and computing all effective critical  vertices.}
\label{al:algo2}
\end{algorithm}\\
Actually finding all effective critical vertices, is the most important work for computing visibility area of $ \mathcal{P} $ from viewpoint $ q $. Now, if we face $ i $-th critical vertex, we find whether it is in visibility area from $ q $ by looking at $ w_i\in W $.
\newpage
For computing $ W $, we do two procedures:
\begin{itemize}
  \item Finding critical vertices of $ \mathcal{P} $ that are not effective and providing remained critical-min and critical-max vertices in ascending order, separately in two sequence, in the linear-time using $O(c+\log n) $ bits space.
  \item Merge two above sequences and computing all effective critical vertices in the linear-time, using same $O(c+\log n) $ bits space.
\end{itemize}
Now, we have the following theorem:
\begin{theorem}
Given a simple polygon $ \mathcal{P} $ in a read-only array $ \mathcal{A} $ of $ n $ vertices $ O(\log n)$ bits each and a viewpoint $ q $ inside $ \mathcal{P} $. Using $ O(c+\log n) $ bits space (equal to $ O(c/\log n)$ additional variables of workspace), there is an algorithm which computes all visible critical vertices in the linear-time, where $ c $ is the number of all critical vertices. See Algorithm~\ref{al:algo1},\ref{al:algo2} and Video~\ref{vi:video1}~\ref{vi:video2} to illumination.
\end{theorem}

\section{An algorithm for computing visibility area of simple polygons}
\begin{figure}[t]
    \centering
    \includegraphics[width=0.8\textwidth]{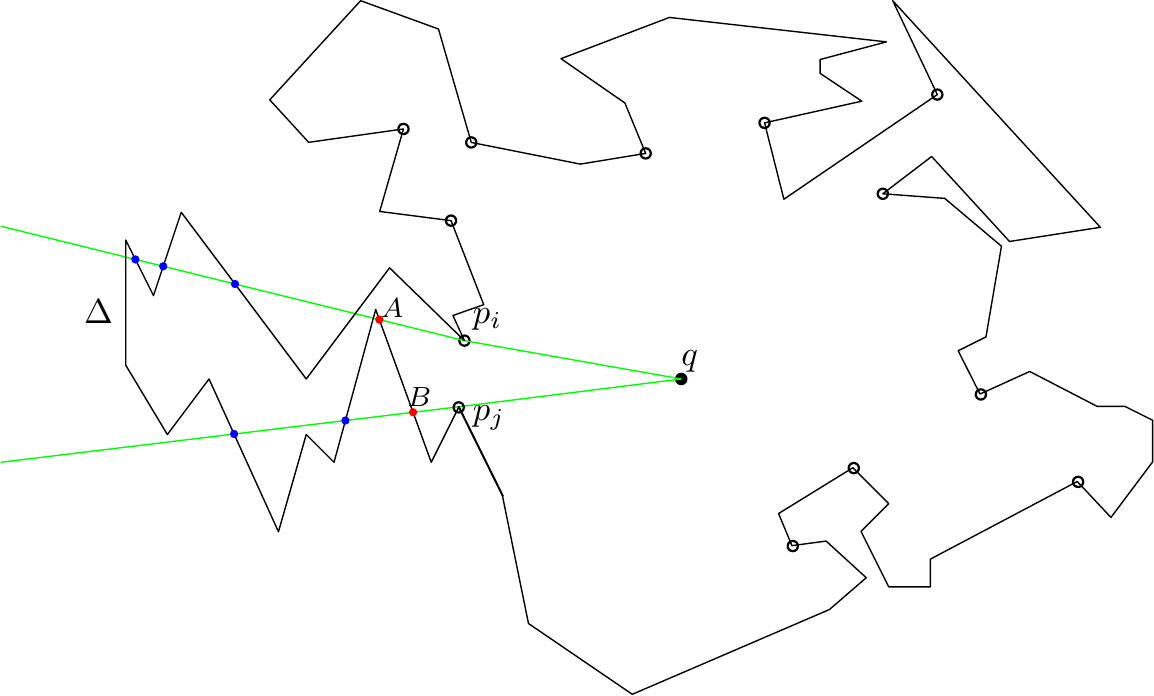}
    \caption{The circles are effective critical vertices, the red and blue points are intersections of $ qp_i $, $ qp_j $ with the chain $ \Delta $, and red points are vertices of visibility area of $ \Delta $ from $ q $.}
    \label{fi:fig3}
\end{figure}

In this section, we assume that all $ \overline{c} $ visible (effective) critical vertices of $ \mathcal{P} $ are reported in $ W $. These vertices divide the boundary of $ \mathcal{P} $ into $ \overline{c} $ separate chains. Let $ p_i $ and $ p_j $ are two consecutive effective critical vertices and the chain that is located between them is called $ \Delta $. If we provide a linear-time algorithm to computing visibility area of $ \Delta $ from $ q $ (linear-time in the number of $ \Delta $), then computing visibility area of $ \mathcal{P} $ is computable in $ O(n) $-time. See Figure~\ref{fi:fig3}.
\newpage
So, we have assumed that $ p_i $ and $ p_j $ are the start and end points of $ \Delta $ (i.e. let $ \theta_i<\theta_j $), respectively. While $ p_i $ is not a critical-max and $ p_j $ is not a critical-min, we report $ \Delta $ as the visibility area. If $ p_i $ is a critical-max, then we compute all the intersections between line passes from $ qp_i $ and every edge of $ \Delta $, and denote the nearest one by $ A $, else we denote $ p_i $ by $ A $. Similarly, if $ p_j $ is a critical-min, then we compute all the intersections between the line passes from $ qp_j $ and every edge of $ \Delta $, and denote the nearest one by $ B $, else we denote $ p_j $ by $ B $. After that, we report the boundary of visibility area as:
\begin{center}
\textit{$ p_i $, $ A $, the vertices between $ A $ and $ B $ on chain $ \Delta $, $ B $, $ p_j $.}
\end{center}
Now, we have our first main theorem in this section:
\begin{theorem}
Given a simple chain $ \Delta $ in a read-only array $ \mathcal{A} $ of $ m $ vertices $ O(\log n) $ bits each and a viewpoint $ q $ in the plane. Using $ O(\log n) $ bits space (equal to $ O(1)$ additional variables of workspace). If $ \Delta $ has only two visible critical vertices, there is an algorithm which computes the visibility area of $ \Delta $ in the linear-time. See Algorithm~\ref{al:algo3} in Appendix~\ref{ap:appendix2} and Video~\ref{vi:video3}.
\end{theorem}
We run this algorithm for every two critical vertices marked in $ W $. By doing it, the sequence of visibility area of $ \mathcal{P} $ will written in $ \mathcal{A'} $, respectively i.e. we traverse on the boundary of $ \mathcal{P} $ counterclockwise until finding two consecutive visible critical vertices, using checking the bits of $ W $. For example, let $ \alpha_1 $ and $ \alpha_2 $ are these two, so, we compute the visibility area of the chain between them by traverse the chain vertices. After that we continue traversing from $ \alpha_2 $ to find next consecutive visible critical vertex, for example $ \alpha_3 $, we compute the visibility area of new chain between $ \alpha_2  $ and $ \alpha_3 $ by traverse the chain vertices. This processing continues until we find all visibility area of $ \mathcal{P} $. Now, we have the following theorem:
\begin{theorem}
Given a simple polygon $ \mathcal{P} $ with $ n $ vertices in a read-only array and a viewpoint $ q $ inside $ \mathcal{P} $. Using $ O(c/\log n) $ variables or $ O(c+\log n) $ bits of working space, there is an algorithm which computes visibility area of $ \mathcal{P} $ inside $ q $ in $ O(n) $ time. Where $ c $ is the number of critical vertices.
\end{theorem}
\begin{algorithm}[h]
    \KwIn{Vertices of simple polygon $ \mathcal{P} $ in read-only array $ \mathcal{A} $, a view point $ q $ in $ \mathcal{P} $}
    \KwOut{Vertices of visibility area of $ \mathcal{P} $ from $ q $ in write-only array $ \mathcal{A'} $, ordered counterclockwise.}
    \ForEach{two consequtive effective critical vertices $ p,p' $}{
    Algorithm~\ref{al:algo3}\;
    }
\caption{Computing visibility area of simple polygons.}
\label{al:algo4}
\end{algorithm}

\newpage
\section{Conclusion}
We studied the problem of computing visibility of a viewpoint inside a simple polygon. We used a sub-linear workspace for our presented algorithm and kept the time-complexity linear-time and optimal, but our algorithm is not the first one. We read our input from a read-only array and wrote our output in a write-only array. For working space, we used $ c $ bits where $ c $ was the number of critical vertices and a constant number of variables. The working space was both writable and readable. Every used variable has $ O(\log n) $ bits of space. This number of bits ($ O(\log n) $ bits) is required to store one index of the vertices, for example. So, the space-complexity of our algorithm is $ (c/\log n) $ variables. Our aim was keeping time optimally and reducing space complexity. It will excepted to solve this problem with constant variables of workspace in the future work.

\bibliographystyle{plain}
\bibliography{bibfile}
\appendix
\newpage
\section{Appendices}

\subsection{List of Videos}
\label{ap:appendix1}
\begin{video}
\label{vi:video1}
The video is available at~\href{https://www.dropbox.com/s/1flvb04zscdwug2/Algorithm%201.mp4?dl=0}{\textit{\underline{DropBox}}}.
\end{video}
\begin{video}
\label{vi:video2}
The video is available at~\href{https://www.dropbox.com/s/wiykwcas78zi823/Algorithm%202.mp4?dl=0}{\textit{\underline{DropBox}}}.
\end{video}
\begin{video}
\label{vi:video3}
The video is available at~\href{https://www.dropbox.com/s/j0214sowyalhh63/Algorithm%204.mp4?dl=0}{\textit{\underline{DropBox}}}.
\end{video}
\subsection{Additional Algorithms}
\label{ap:appendix2}
\begin{algorithm}[h]
    \KwIn{Vertices of chain $ \Delta $ in random-access array $ \mathcal{A} $, a view point $ q $. The start point $ p $ and end point $ p' $ of $ \Delta $ are its only two visible critical vertices.}
    \KwOut{Write-only array $ \mathcal{A'} $ containing the vertices of visibilty area of $ \Delta $ from viewpoint $ q $, in counterclockwise order.}
    $ V\leftarrow p $; $ V'\leftarrow$ next-vertex\;
    $ min_1\leftarrow \abs{qp} $; $ min_2\leftarrow \abs{qp'} $\;
    \While{$V'\neq p'$}{
    $ x \leftarrow $the intersection between line $ qp $ and the edge $ VV' $(if available)\;
    $ y \leftarrow $the intersection between line $ qp' $ and the edge $ VV' $(if available)\;
    $ min_1\leftarrow \min\{min_1,\abs{qx}\} $\;
    $ min_2\leftarrow \min\{min_2,\abs{qy}\} $\;
    $ V\leftarrow V' $; $ V'\leftarrow $next-vertex\;
    }
    $ V\leftarrow p $; $ V'\leftarrow$ next-vertex\;
    write($ p $) in $ \mathcal{A'} $\;
    \While{the distance from the intersection between $ VV' $ and $ qp $ to $ q $ is not $ min_1 $}{
    $ V\leftarrow V' $; $ V'\leftarrow $next-vertex\;
    }
    write(the intersection between $ VV' $ and $ qp $) in $ \mathcal{A'} $\;
    \While{the distance from the intersection between $ VV' $ and $ qp' $ to $ q $ is not $ min_2 $}{
    write($ V' $) in $ \mathcal{A'} $\;
    $ V\leftarrow V' $; $ V'\leftarrow $next-vertex\;
    }
    write(the intersection between $ VV' $ and $ qp' $) in $ \mathcal{A'} $\;
    write($ p' $) in $ \mathcal{A'} $\;    
\caption{Computing visibility area of chains.}
\label{al:algo3}
\end{algorithm}

\end{document}